\documentclass[12pt]{iopart}
\pdfoutput=1
\usepackage{graphics}
\usepackage{iopams}
\usepackage{amssymb}
\usepackage{amsfonts}
\usepackage{epsfig}
\usepackage{graphicx}
\usepackage{color}
\pdfoutput=1
\usepackage{subfigure}

\begin{document}
\title{Simulating the initial growth of a deposit from colloidal suspensions}
\author{T. J. Oliveira${}^{1}$ and F. D. A. Aar\~ao Reis${}^{2}$}

\address{${}^{1}$ Departamento de F\'isica, Universidade Federal de Vi\c cosa, 36570-900,
Vi\c cosa, MG, Brazil}
\address{${}^{2}$ Instituto de F\'\i sica, Universidade Federal Fluminense, Avenida Litor\^anea s/n,
24210-340 Niter\'oi, RJ, Brazil}

\ead{tiago@ufv.br, reis@if.uff.br}

\date{\today}

\begin{abstract}

We study the short time properties of a two-dimensional film growth model in which incident particles
execute advective-diffusive motion with a vertical step followed by $D$ horizontal steps. 
The model represents some features of the deposition of anisotropic colloidal particles of the
experiment in Phys. Rev. Lett. {\bf 110}, 035501 (2013), in which wandering particles are attracted
to particle-rich regions in the deposit. Height profiles changing from rough to columnar structure
are observed as $D$ increases from $0$ (ballistic deposition) to $8$, with striking similarity to
the experimental ones. The effective growth exponents matches the experimental estimates and
the scaling of those exponents on $D$ show a remarkable effect of the range of the
particle-deposit interaction. The nearly ellipsoidal shape of colloidal particles
is represented for the calculation of roughness exponents in conditions that parallel the
experimental ones, giving a range of estimates that also include the experimental values.
The effective dynamic exponents calculated from the autocorrelation function are shown to be
suitable to decide between a true dynamic scaling or transient behavior, particularly because
the latter leads to deviations in an exponent relation. These results are consistent with arguments 
on short time unstable (columnar) growth of Phys. Rev. Lett {\bf 111}, 209601 (2013), indicating that 
critical quenched KPZ dynamics does not explain that colloidal particle deposition problem.

\end{abstract}


\maketitle

\section{Introduction}
\label{intro}

The scaling properties of growing interfaces attracted much interest in the last decades
due to the large number of problems of technological and fundamental interest in which
the physical and chemical properties are affected by interface morphology
\cite{barabasi,ohring}. This scenario motivated the proposal of models based on stochastic equations,
which represent the main symmetries of the growth processes and define universality classes,
and the development of atomistic models that capture the details of those processes and are
frequently suitable for particular applications. Among the hydrodynamic models, the
Kardar-Parisi-Zhang (KPZ) equation \cite{kpz} deserves special interest due to the
large number of possible applications and the recent progress in its solution
\cite{sasamoto,amir,calabrese,imamura}.

However, there are frequent debates on the universality class of growth processes
in which different scaling exponents are measured when the growth conditions are changed
(e. g. changes in temperature, reactant flux, or electric potential)
\cite{ebothe,kleinke,naraNano2007,sukarno,huo,lafouresse,placidi}. Several stochastic equations
and atomistic models actually show variations of
scaling exponents as the growth conditions change. In many cases, these features are related to
crossovers between different dominant dynamics at different time and length scales
\cite{majaniemi,krug1999,GGG,NT,tiagoewkpz,rdcor,albano}.
Many of those models show crossover to KPZ scaling.

In a recent work, Yunker et al. \cite{yunker} studied the deposition of colloidal
particles in the water-air interface of evaporating drops, showing that
scaling exponents of height fluctuations and the height distributions changed
with the aspect ratio $\epsilon$ of the particles. As $\epsilon$ increased, those changes were explained as
transitions from uncorrelated (Poisson-like) deposition to normal KPZ
scaling and then to critical KPZ scaling with quenched disorder (QKPZ) \cite{kpzdisordered} in
$1+1$ dimensions. The main physical change in the system as $\epsilon$ increases seemed to be
the onset of a kind of ``Matthew effect'' \cite{merton}, in which wandering particles are attracted 
to rich-particle regions in deposit. It explains the transition from Poisson to ballistic (KPZ) growth
when the particle shape changes from spherical (no interaction) to slightly elliptical (small attraction).
However, Nicoli et al. \cite{nicoli} showed that a morphological instability (and consequent anomalous
roughening) may be the origin of those exponent changes, so that the claim of QKPZ scaling is not necessary
(although Yunker et al. \cite{yunkerreply} argue that the initial roughness of the contact line may
be the source of that disorder).

This scenario motivates the present study of a deposition model of diffusing particles with a drift, which
is an extension of a model proposed by Rodr\'{\i}guez-P\'erez, Castillo and Antoranz \cite{perez2005},
hereafter called RCA model. 
In this model, the mechanisms of particle advection and diffusion before aggregation is consistent with
the above mentioned experiment for large $\epsilon$ \cite{yunker}, since the diffusive flow leads to
preferential sticking at prominent regions of the deposit. The growth of large and thick deposits with
the RCA model showed that it is in the KPZ class in $2+1$ dimensions \cite{perez2005,perez2007}, and here
we show that the extended model is also in that class in $1+1$ dimensions.
However, the main interest of the present work is to study the
scaling of height fluctuations of deposits with small average thicknesses, of the same
order of magnitude of the thicknesses of colloidal particle deposits of Ref. \protect\cite{yunker}.
The effect of particle shape on the surface morphology is analyzed using the intra-grain modeling of
Ref. \protect\cite{grainshape}.

For small values of the advection-to-diffusion rate (Peclet number), our model provides interface
profiles with columnar structures at short times, which leads to estimates of (effective) growth exponents
that significantly exceed both the KPZ and the uncorrelated deposition values ($1/3$ and $1/2$).
After representing the ellipsoidal particle shapes at the surface, the local roughness exponent
attains large values in nearly two decades of window size.
The interface profiles are similar to the colloidal particle deposits of
Ref. \protect\cite{yunker}, for average heights of the same order, and several values of effective
exponents are in the ranges reported in that work.
Scaling arguments explain the dependence of the growth exponent on the lateral diffusion coefficient
of the model, showing that small changes in the range of attraction to the deposit
may lead to significant changes in that exponent.
Thus, a description of that experiment without accounting for critical
disorder mechanisms seems to be possible, in agreement with Ref. \protect\cite{nicoli}.
The study of the scaling of the zeros and minima of the
autocorrelation function is proposed as an additional test for the universality class in these problems.
Our model is also shown to have asymptotic KPZ scaling, which raises the
possibility that the colloidal particle deposition is also in the KPZ class for any $\epsilon>1$, but with    
strong scaling corrections.

The rest of this works is organized as follows. In Sec. \ref{models} the deposition model is presented in detail 
as well as the definition of all quantities analyzed in this work. Sec. \ref{interfacegrowth} shows
results for deposits and height profiles and discuss the variation in the growth exponents.
In Sections \ref{dynamic} and \ref{roughness}, results for dynamic and local roughness exponents are shown,
respectively. In Sec. \ref{kpz}, we provide numerical evidence that the model is asymptotically in the KPZ class.
Our final conclusions are presented in Sec. \ref{conclusion}.

\section{Model, simulation and quantities of interest}
\label{models}

The model is defined in a square lattice with an initial planar substrate at $y=0$.
Each site may be empty or occupied by a particle of linear size $d$.
Each deposition step begins with a new particle being left at a random position
above the deposit. This particle executes steps in a ballistic-diffusive sequence, which consists
of one vertical down step followed by $D$ horizontal random steps. When this particle reaches
a site with a nearest neighbor occupied site or a substrate site, it permanently aggregates there.
The time unity is defined as the time of deposition of $L$ particles.

This model is slightly different from the RCA model because the latter allows the random
steps to be performed in the horizontal or in the vertical direction.
For more details on the physico-chemical motivation of the RCA model, we suggest the reader to
consult Ref. \protect\cite{perez2005}. Similar models with multiparticle flux were proposed
in Refs. \protect\cite{castro1998,castro2000} and a recent extension in
Ref. \protect\cite{galindo}.

\begin{figure}[t]
\centering
\includegraphics[width=11.5cm]{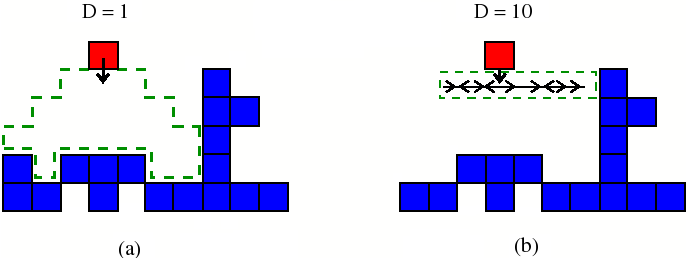}
\caption{(Color online) Flow of a (red) particle towards the (blue) deposit with (a) $D=1$ and
(b) $D=10$. The region that the particle in (a) can reach after several vertical and horizontal
steps is surrounded by the (green) dashed line, showing that aggregation will occur at the valley of
the deposit. The region scanned by the particle in (b) after a single vertical step
is also surrounded, which leads to aggregation to the protruding part of the deposit.}
\label{fig1}
\end{figure}

Diffusion-limited aggregation (DLA) \cite{witten} is a widely used model for electrochemical deposition
\cite{castro1998,castro2000,silvio2005,nicoli2009}. Indeed, an electrodeposition model proposed
in Ref. \protect\cite{hill} has stochastic rules similar to the present model.
Particle diffusion facilitates the aggregation at the protuberant parts of the deposit,
thus representing the attraction of particles to those regions.
On the other hand, the directed particle flux is characteristic of ballistic deposition (BD)
\cite{vold,barabasi} (the RCA model interpolates BD for $D=0$ and DLA for $D\to\infty$).

Figs. \ref{fig1}a and \ref{fig1}b illustrate the consequence of the horizontal diffusive motion when a
particle incide near a protruding part of a deposit. For small $D$ (Fig. \ref{fig1}a),
the incoming particle is not able to stick to that surface feature because it rapidly flows towards the
substrate. 
However, for large $D$ (Fig. \ref{fig1}b), it samples a wide horizontal region after each vertical step,
thus it is possible to attach to the protruding point.

In the colloidal particle deposition of Ref.\protect\cite{yunker}, an attraction of elongated
(large aspect ratio $\epsilon$)
particles to the protruding points of the deposit is observed. This is confirmed by the preferential
sticking to the hills of the slightly rough contact line in the initial steps of the deposition
\cite{yunkerreply}. For this reason, the application of our model to that problem requires
larger values of $D$ to be associated to larger values of $\epsilon$. In this
case, $D$ should not be interpreted as a diffusion coefficient in solution. Instead,
the scanned length proportional to $D^{1/2}$ (Fig. \ref{fig1}b) may be related to the range of the
attractive particle-deposit interaction.

Simulations of the model were performed for several values of $D$ between $D=0$ (BD) and $D=8$,
in a substrate with $4096$ sites, which is sufficiently large to avoid finite-size effects.
The time unit is set as that necessary for the deposition of one layer of particles
($4096$ particles); since the deposits are porous, the average thickness after one time unit
exceeds the particle size $d$.

The global roughness of the surface of a deposit at time $t$ is defined as the
rms fluctuation of the height $h$ around its average position $\overline{h}$:
\begin{equation}
w(t)\equiv { \left< \overline{ {\left( h - \overline{h}\right) }^2 }^{1/2} \right> } ,
\label{defw}
\end{equation}
where the overbars indicate spatial averages and the angular brackets indicate configurational averages.
The local roughness $w_{loc}(l,t)$ is defined as the rms height fluctuation averaged inside
a box of linear size $l$ that  slides along the surface, parallel to the substrate.

The autocorrelation function is defined as
\begin{equation}
\Gamma\left( r,t \right)\equiv \langle \tilde{h}\left( x,t\right) \tilde{h}\left( x+r,t\right) \rangle ,
\label{defgamma}
\end{equation}
where $\tilde{h} \equiv h - \overline{h}$ and the brackets represent an
average over different positions $x$ along the substrate.

In sufficiently large substrates (in which finite-size effects are negligible),
the global roughness scales as 
\begin{equation}
w\sim t^\beta ,
\label{defbeta}
\end{equation}
where $\beta$ is called roughness exponent.
For box sizes smaller than the lateral correlation length, the local roughness
of a system with normal scaling depends on $l$ as
\begin{equation}
w_{loc}\sim l^\alpha ,
\label{defalpha}
\end{equation}
where $\alpha$ is the roughness exponent [in systems with anomalous scaling \cite{ramasco},
$\alpha$ is replaced by the local roughness exponent in Eq. (\ref{defalpha}), which
differs from the global roughness exponent].

The first zero ($r_0$) or the first minimum ($r_m$) of the autocorrelation function
$\Gamma\left( r,t \right)$ gives an estimate of the lateral correlation length of the surface
at time $t$. They are expected to scale as
\begin{equation}
r_0, r_m \sim t^{1/z} ,
\label{defz}
\end{equation}
where $z=\alpha /\beta$ is the dynamic exponent.

Since the colloidal particle deposition of Ref.\protect\cite{yunker} is a motivation for this
study, it is important to account for the order of magnitude of the quantities involved there.
The average diameter of spherical particles is $\langle d\rangle =1.3 \mu m$,
and elongated particles have similar characteristic size.
The maximal average height of the deposits is near $30 \mu m$, which amounts to $23\langle d\rangle$.
This justifies the study of short time features of the model, since that thickness corresponds to
deposition of nearly $23$ complete layers in a compact film, but much less in a porous film.

A more subtle point is that the roughness exponent was measured
in box sizes ranging from $l_{min}\approx 0.1\mu m$ to $l_{max}\approx 10 \mu m$  in Ref.\protect\cite{yunker},
(the local roughness is measured until saturation,
in sizes up to $l\approx 200\mu m$ or $300 \mu m$).
These sizes range from $l_{min}\approx 0.08d$ to $l_{max}\approx 8d$, thus the scaling exponents are
probing height fluctuations inside a single particle (intra-particle features) as well as
height fluctuations of sets with several particles. This justifies the use of models
describing intra-particle features along the lines of Refs. \protect\cite{graos,grainshape}.

\section{Interface profiles and growth exponents}
\label{interfacegrowth}

\begin{figure*}[t]
\includegraphics*[width=5.3cm]{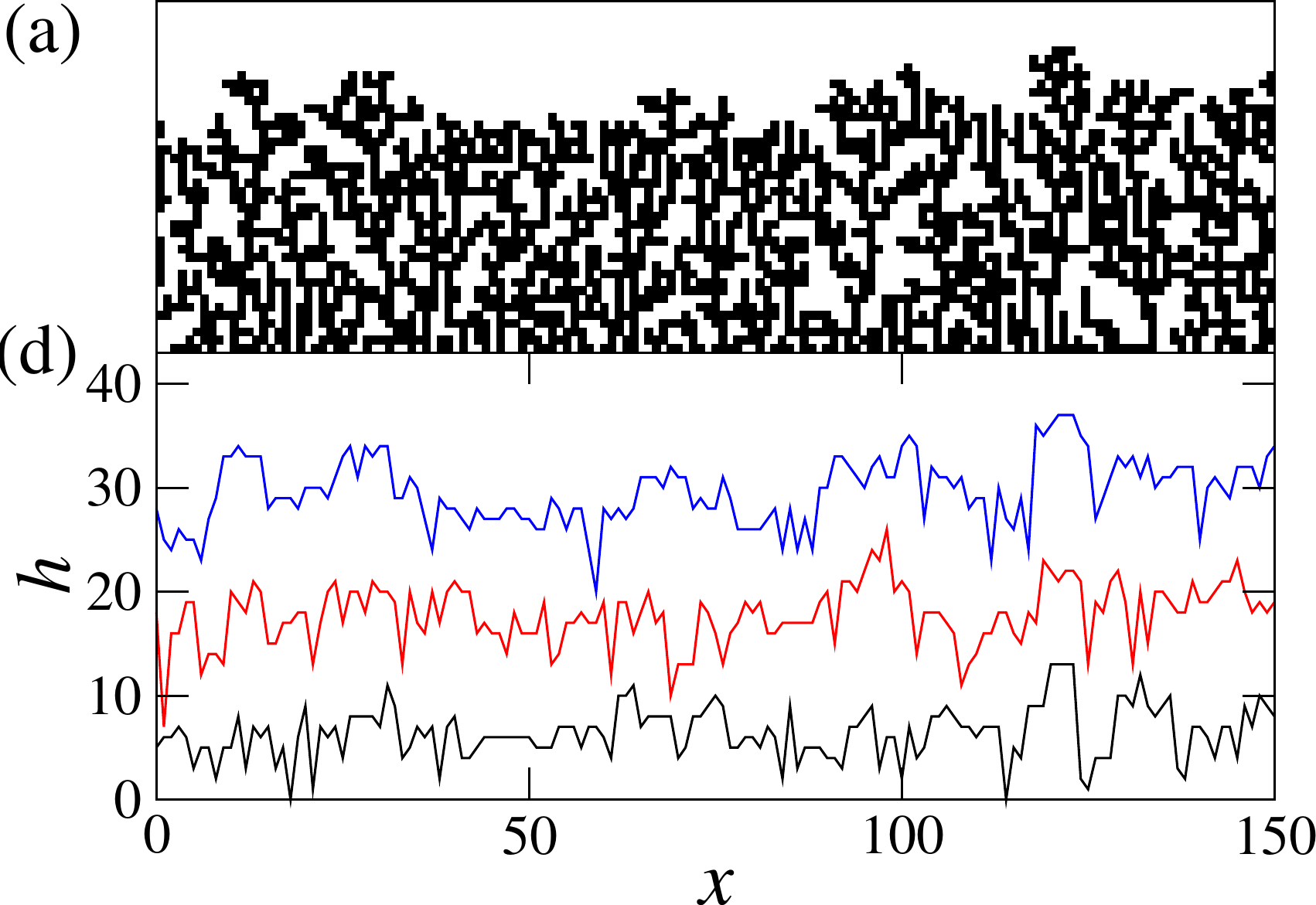}
\includegraphics*[width=5.3cm]{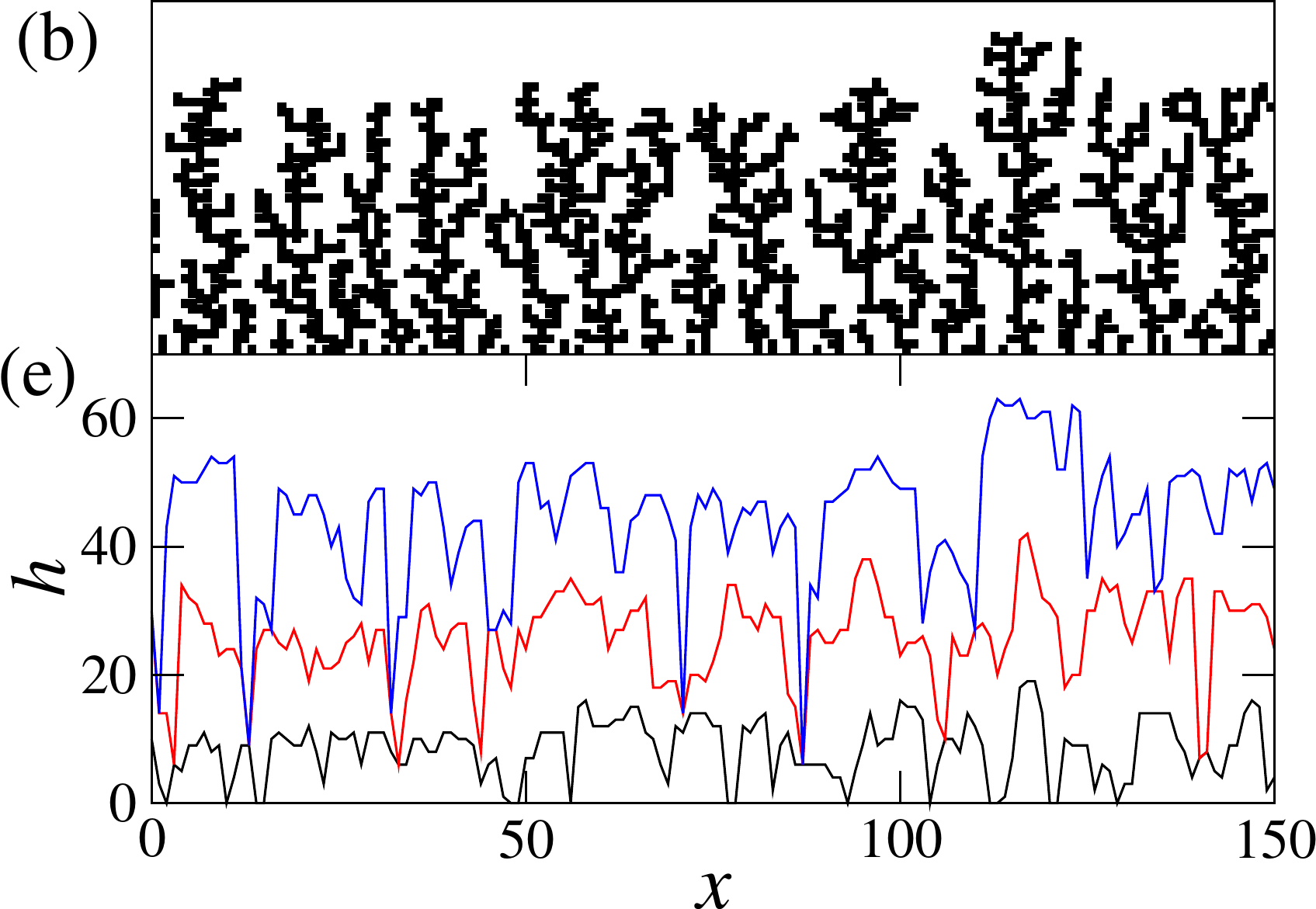}
\includegraphics*[width=5.3cm]{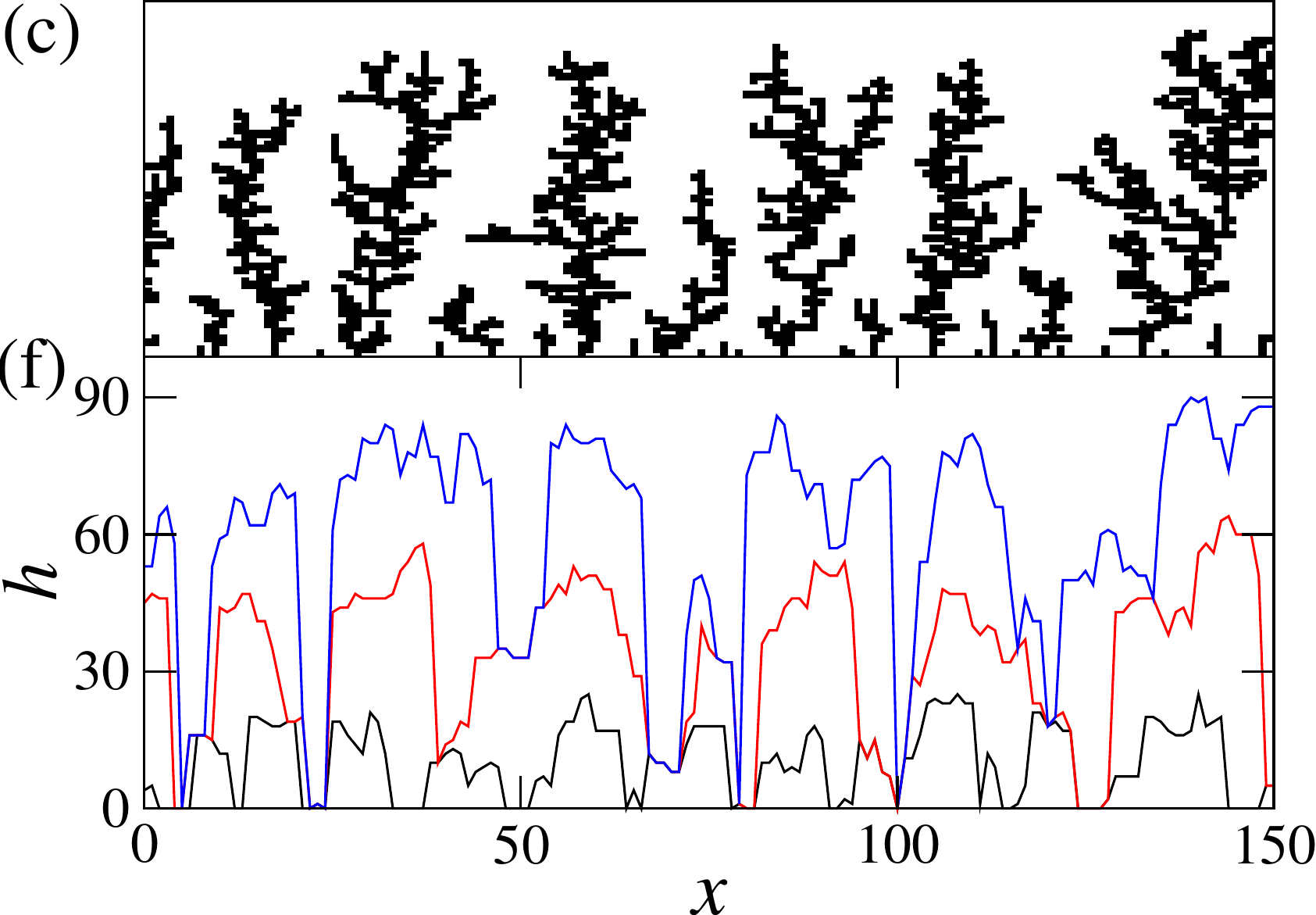}
\caption{(Color online) Typical deposits (top) and height profiles (bottom) of the RCA model after
deposition of 4 (black), 10 (red), and 16 (blue) particle layers, for: (a) $D=0$, (b) $D=2$, and (c) $D=8$.
$h$ and $x$ are measured in units of the particle size $d$.}
\label{fig2}
\end{figure*}

Figs. \ref{fig2}a-c show sections of the deposits generated by the model with $D=0$ (BD), $D=2$,
and $D=8$, after deposition of $16$ layers of particles. The increase of $D$ is
responsible for the decrease of the film density, as observed in Ref.
\protect\cite{perez2005}. For this reason, the same deposition time leads to very
different average heights, which range from $30d$ to $61d$ in Figs. \ref{fig2}a-c.
Figs. \ref{fig2}d-f show the interface profiles at three different times for the same values
of $D$ of Figs. \ref{fig2}a-c (the interfaces for the longest time correspond to the deposits shown
in Figs. \ref{fig2}a-c). The interface height at each column $x$ [$h(x)$] is defined as the
largest $y$ value of an aggregated particle at $x$.

The branched structure shown in Figs. \ref{fig2}a-c is similar to that obtained in the RCA model
\cite{perez2005}.
However, a surprising result is the combination of this branched structure of the deposits and
the columnar structure of the corresponding interface profiles (Figs. \ref{fig2}d-f), which parallel the
features of  colloidal particle deposits of Ref. \protect\cite{yunker} with large $\epsilon$.

For large $D$, those columnar structures persist up to relatively long times. 
For this reason, a long crossover to KPZ scaling is expected with large $D$.
Surfaces with columnar features are
characteristic of processes in which the local growth rate is spatially inhomogeneous
(larger at the hills, smaller at the valleys), in contrast with the translational
symmetry of the KPZ and similar stochastic equations \cite{barabasi}.
With columnar structures, the height fluctuations between hills and valleys increase linearly
in time, thus a growth exponent  [Eq. \ref{defbeta}] near $1$ is expected.

\begin{figure}[t]
\centering
\includegraphics[width=8.5cm]{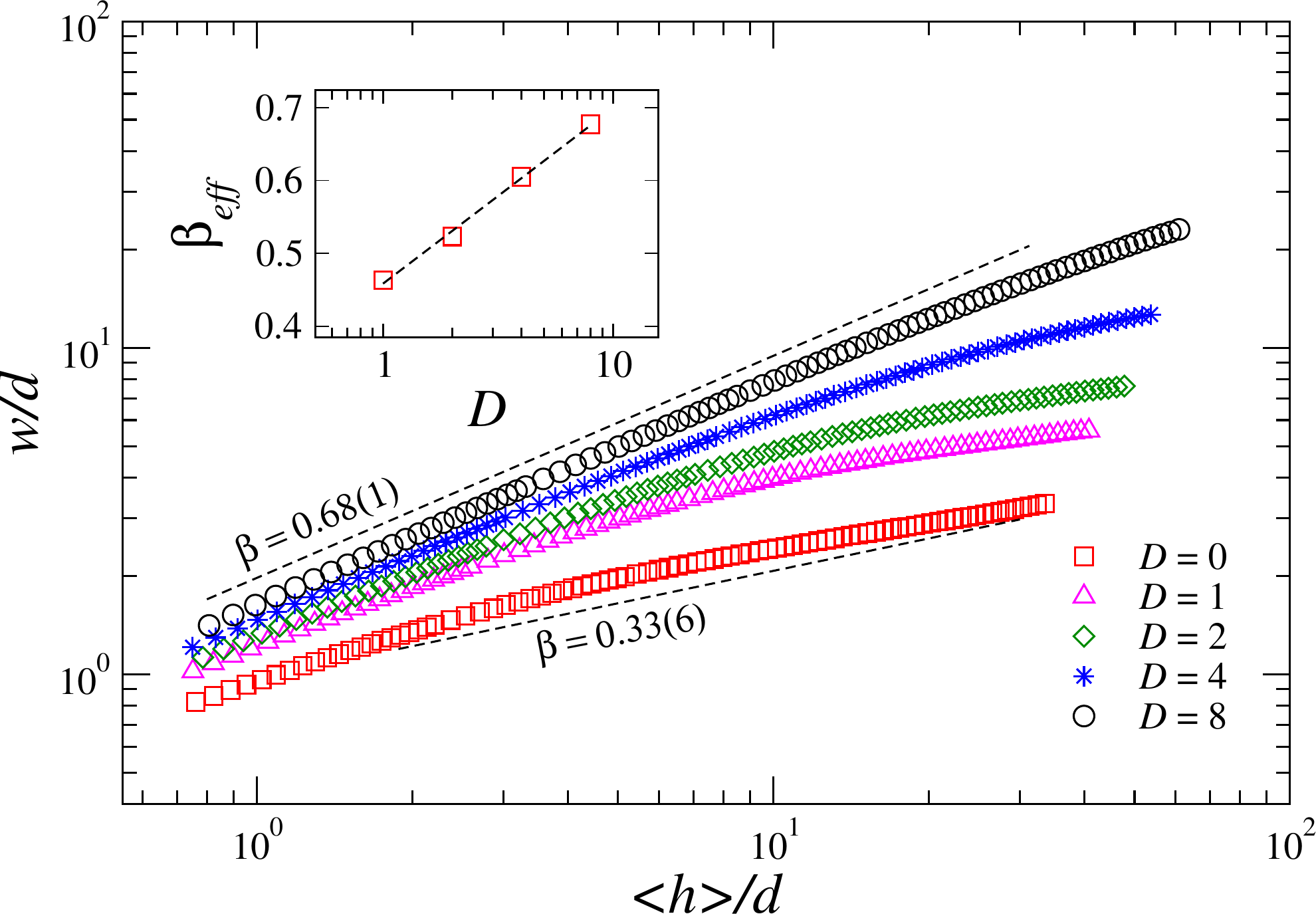}
\caption{(Color online) Interface width $w$ versus average height $\left\langle h \right\rangle$,
rescaled by the particle size $d$, for the RCA model with several values of $D$.
Dashed lines have the slopes indicated in the plot. The inset shows the effective growth exponents
as a function of $D$.}
\label{fig3}
\end{figure}

Fig. \ref{fig3} shows the global roughness of the RCA films as a function of the average height
for several values of $D$. The average height is nearly proportional to
the deposition time $t$, but it is a more interesting quantity for comparison with experiments.
The roughness clearly increases with $D$ for deposits with the same thickness,
in agreement with previous results \cite{perez2005}.

For $D=0$, a short time crossover is clear from the downward curvature of the
$\log{w}\times\log{\langle h\rangle}$ plot, thus the estimate of $\beta$ is highly dependent
on the range of $\langle h\rangle$ chosen to fit. For example, for 
$2\lesssim\langle h\rangle /d\lesssim 30$, we found local slopes continuously decreasing 
from $0.39$ to $0.27$. This range is used to determine the error bar in the estimate of $\beta$.
The average slope is very close to the KPZ 
exponent $\beta_{KPZ}=1/3$ \cite{kpz}, as shown in Fig. \ref{fig3}. However, BD is known to have a 
long crossover to KPZ scaling, so that deviations to smaller values of $\beta$ are expected for 
slightly longer times, as shown in Ref. \protect\cite{balfdaar} and observed here.

For large $D$, reasonable linear fits can be obtained in almost two orders of magnitude
of $\langle h\rangle$, as illustrated in Fig. \ref{fig3}. For $D=8$, the growth exponent
$\beta\approx 0.68$ is measured in that region and is
much larger than the values of KPZ ($\beta=1/3$) and
of uncorrelated deposition ($\beta =0.5$). This is consistent with the columnar growth illustrated in
Fig. \ref{fig2}f. Thus, in applications with small film thicknesses, the scaling of the surface roughness
may provide misleading information on the universality class of the growing system.

Fig. \ref{fig3} considers thicknesses ranging from submonolayer coverage ($\langle h\rangle <d$)
to some tens of the particle size, which is suitable for comparison with  Ref. \protect\cite{yunker}.
The exponents $\beta$ at short times are in good agreement with those
of the colloidal particle deposition of Yunker et al \cite{yunker} for slightly anisotropic
particles ($\epsilon =1.2$, $\beta\approx 0.37$) and for highly anisotropic ones
($\epsilon =3.5$, $\beta\approx 0.68$). However, the values of $\beta$ for the model vary continuously between
these limits as $D$ increases, which differs from the apparent jumps in the experimental values of
$\beta$ (although the existence of a continuous but rapid change in the experimental exponents cannot be discarded).

A simple relation between the effective growth exponent and $D$ can be derived with scaling
arguments, as follows.
For very short times, $w/d\approx 1$ and weakly depends on
$D$, since submonolayer growth dominates, with rare aggregation at the second or third layers.
However, for large thicknesses (e. g. $\langle h\rangle/d\sim {10^2}$),
the lateral diffusion length of incident particles is
$l_{\|}\sim D^{1/2}$ for a small number of vertical steps, which leads to formation of valleys
between the columns (Fig. \ref{fig2}f). The roughness is proportional to the height difference between 
those valleys and columns, which is expected to scale as $l_{\|}^x$, where $x>0$ is some effective
exponent (we are not attempted to assume that this is the KPZ roughness exponent because
this is a transient regime without KPZ scaling). This leads to $w/d\sim D^{x/2}$. The
corresponding slope of the $\log{w}\times\log{\langle h\rangle}$ is of the form
\begin{equation}
\beta_{eff} \approx a+b\log{D} .
\label{betaeff}
\end{equation}
This is confirmed in the inset of Fig. \ref{fig3}, with $a\approx 0.4$ and $b\approx 0.1$.

Note that $\log{D}$ rapidly changes for small $D$, which leads to significant changes
of the growth exponent shown in Fig. 3. The same rapid change occurs in $\log{l_{\|}}$.
This means that small changes in the range of the attractive particle-deposit interaction
may lead to large changes in the growth exponent.
Indeed, transient scaling may be quite sensitive to details of the interactions, even
if asymptotic scaling is the same.

A point that deserves particular attention in the colloidal particle deposition experiment
is the estimate $\beta\approx 0.48$ for spherical particles ($\epsilon =1$),
which was interpreted as a consequence of uncorrelated random deposition in Ref. \protect\cite{yunker}
(a model in which $\beta =1/2$ \cite {barabasi}).
However, the corresponding $\log{w}\times\log{\langle h\rangle}$ plot
had a large number of data points with average height below $1.3 \mu m$
and global roughness smaller than $0.7 \mu m$ (in contrast to the fits
of $\log{w}\times\log{\langle h\rangle}$ plots for other particle shapes, which have most data
points with $\langle h\rangle >1.3 \mu m$).
In the absence of discontinuities in that plot, those average heights may be representative of
the early submonolayer regime, in which the roughness scales similarly to uncorrelated deposition.

Simulations of our model with $\langle h\rangle <d$ help to understand what is expected if
submonolayer data is included in the calculation of the effective exponent $\beta$
from $\log{w}\times\log{\langle h\rangle}$ plots. 
For small $D$, the submonolayer data gives slopes much larger than the slopes for
$\langle h\rangle >d$, revealing a crossover which may eventually parallel the experiments
with spherical particles. On the other hand, for large $D$, the slope with $\langle h\rangle <d$
is approximately the same measured with $\langle h\rangle >d$.
Anyway, performing fits of the experimental data restricted to $\langle h\rangle >1.3 \mu m$
would be necessary to confirm these proposals.

\section{Dynamic exponents}
\label{dynamic}

Broad ranges of dynamic exponents are also obtained for the short-time RCA model
using the autocorrelation function.

\begin{figure}[t]
\includegraphics*[width=8.3cm]{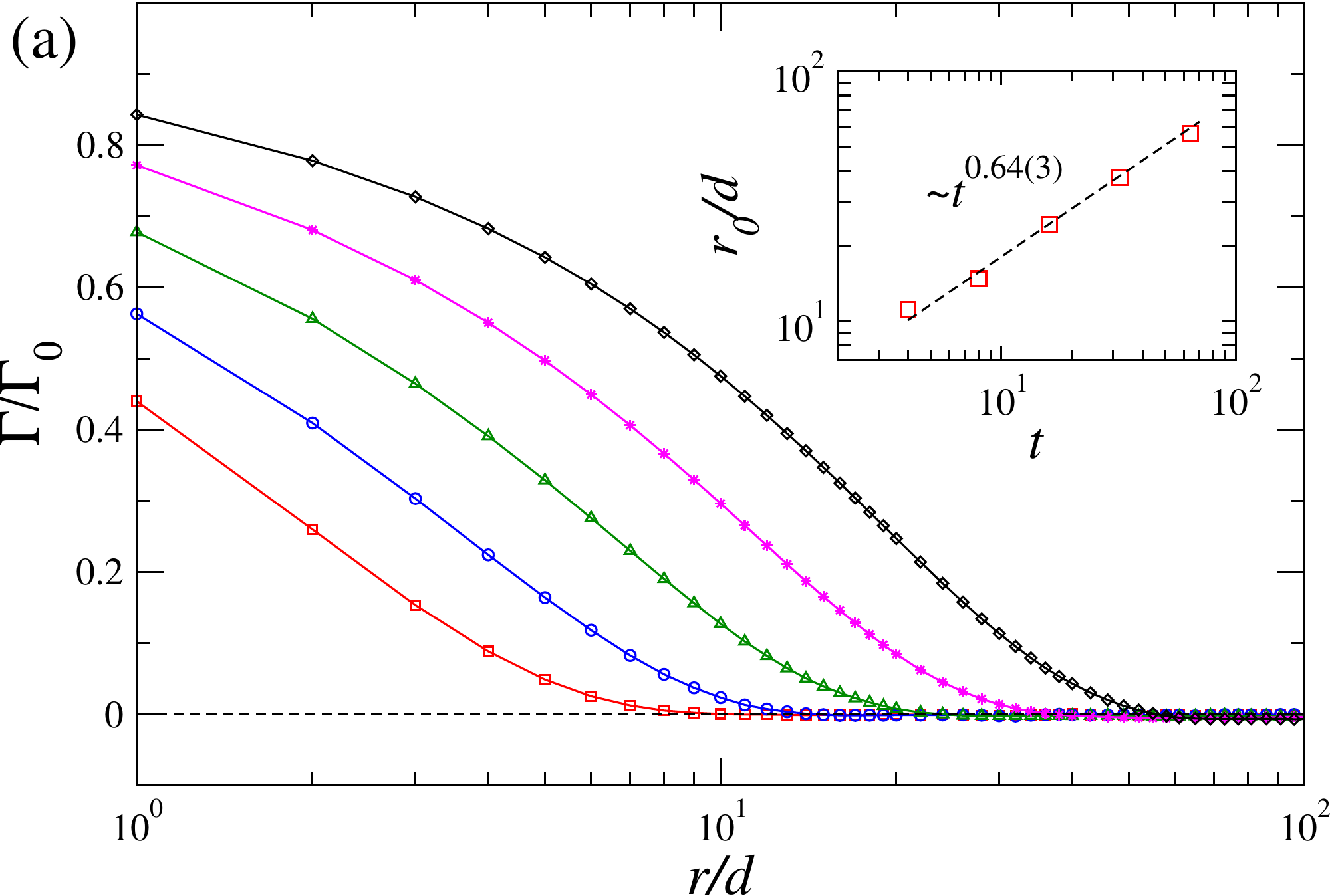}
\includegraphics*[width=8.3cm]{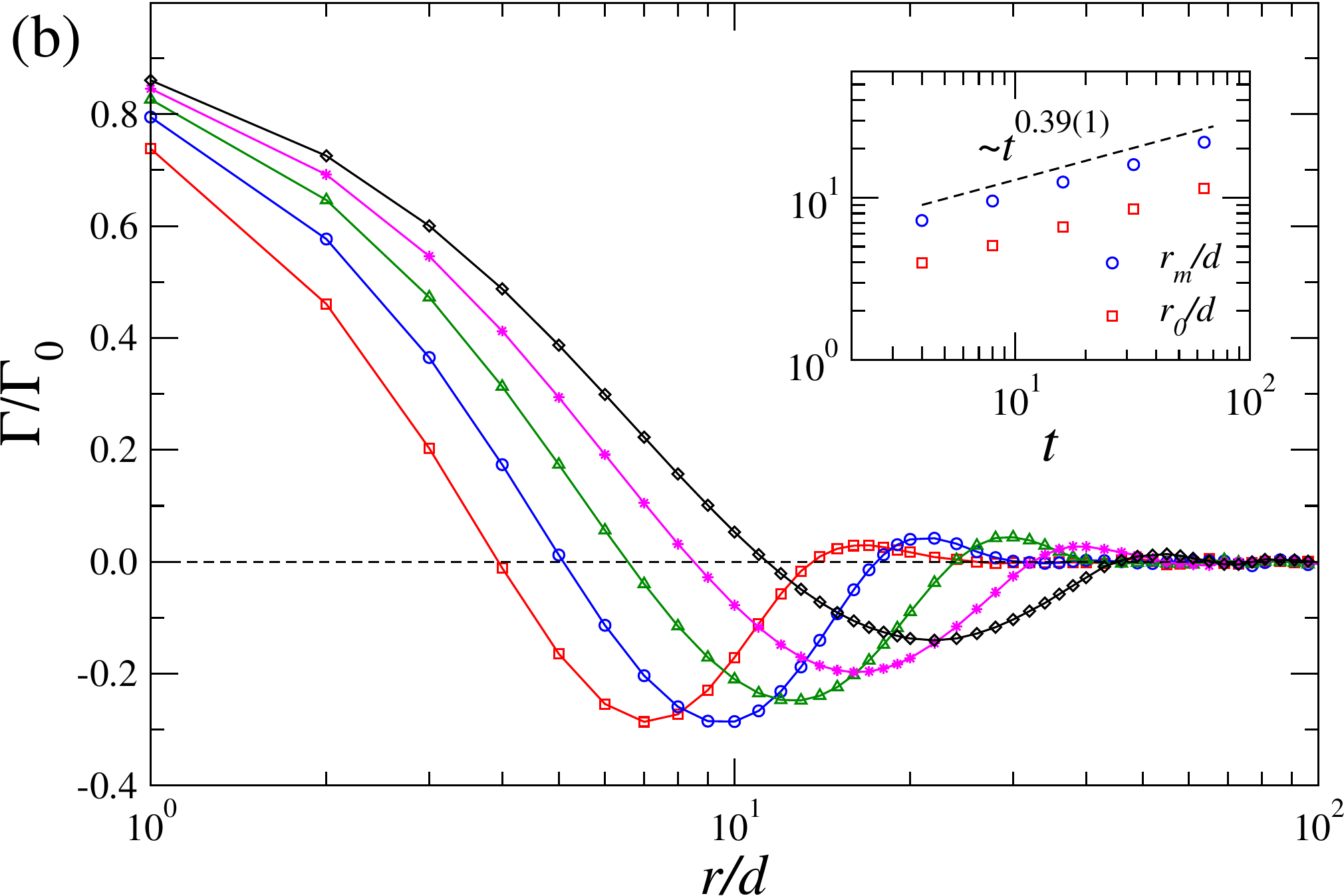}
\caption{(Color online) Normalized autocorrelation function $\Gamma(r)/\Gamma(0)$ versus box size $r$
rescaled by the particle size $d$, for (a) $D=0$ and (b) $D=8$, at deposition times $t=4$ (red squares),
$8$ (blue circles), $16$ (green triangles), $32$ (magenta stars), and $64$ (black diamonds). Times
are measured in number of deposited monolayers.}
\label{fig4}
\end{figure}

Figs. \ref{fig4}a and \ref{fig4}b show the normalized autocorrelation function as a function of the
distance $r$ for $D=0$ and $D=8$, respectively, considering various average thicknesses. 
For $D=0$, the minimum $r_m$ of $\Gamma\left( r,t \right)$ is difficult to be identified, thus
estimates of $z$ are calculated only from the scaling of $r_0$. For large values of $D$,
$\Gamma\left( r,t \right)$ has a oscillatory structure with a clear minimum (Fig. \ref{fig4}b), which is a
consequence of the columnar interface morphology (see also Ref. \protect\cite{siniscalco}).

The insets of Figs. \ref{fig4}a and \ref{fig4}b show the scaling of $r_0$ and $r_m$ for $D=0$ and $D=8$, respectively.
The estimates of the dynamic exponent are $z=1.56\pm 0.07$ and $z=2.56\pm 0.06$,
respectively. The first estimate agrees with the KPZ value $1.5$, but the second one
is very far from that value. Again, a failure of KPZ scaling is observed for short growth times
and for large lateral mobility of incident particles (large $D$).

The large effective dynamic exponent obtained for $D=8$, as a consequence of the columnar morphology, is
very different from the dynamic exponent $z=1$ of QKPZ scaling.
Thus, the particular features of the plots of the autocorrelation function and the estimates of
dynamic exponents strongly suggest that these quantities should be studied in the colloidal particle 
deposits of Yunker et al \cite{yunker} and related applications in order to understand their growth
dynamics.

Comparison of Figs. \ref{fig4}a and \ref{fig4}b also shows that the apparent correlation length
is much larger for $D=0$ (BD). This occurs because the columnar structure for large $D$
has lateral fluctuations determined by that parameter (as $l_{\|}\sim D^{1/2}$), which is not a KPZ feature.
On the other hand, for $D=0$, the fluctuations are determined by lateral propagation of correlations
of KPZ type. Due to this complex interplay of lateral correlation mechanisms, no simple relation between the
effective dynamical exponent $z$ and the parameter $D$ can be found, in contrast to that for the
growth exponent [Eq. (\ref{betaeff})].

\section{Roughness exponents}
\label{roughness}

In order to study the scaling of the local roughness, we will consider the models proposed in
Ref. \protect\cite{grainshape} to account for the shape of the aggregated particles.

\begin{figure}[t]
\centering
\includegraphics[width=8.5cm]{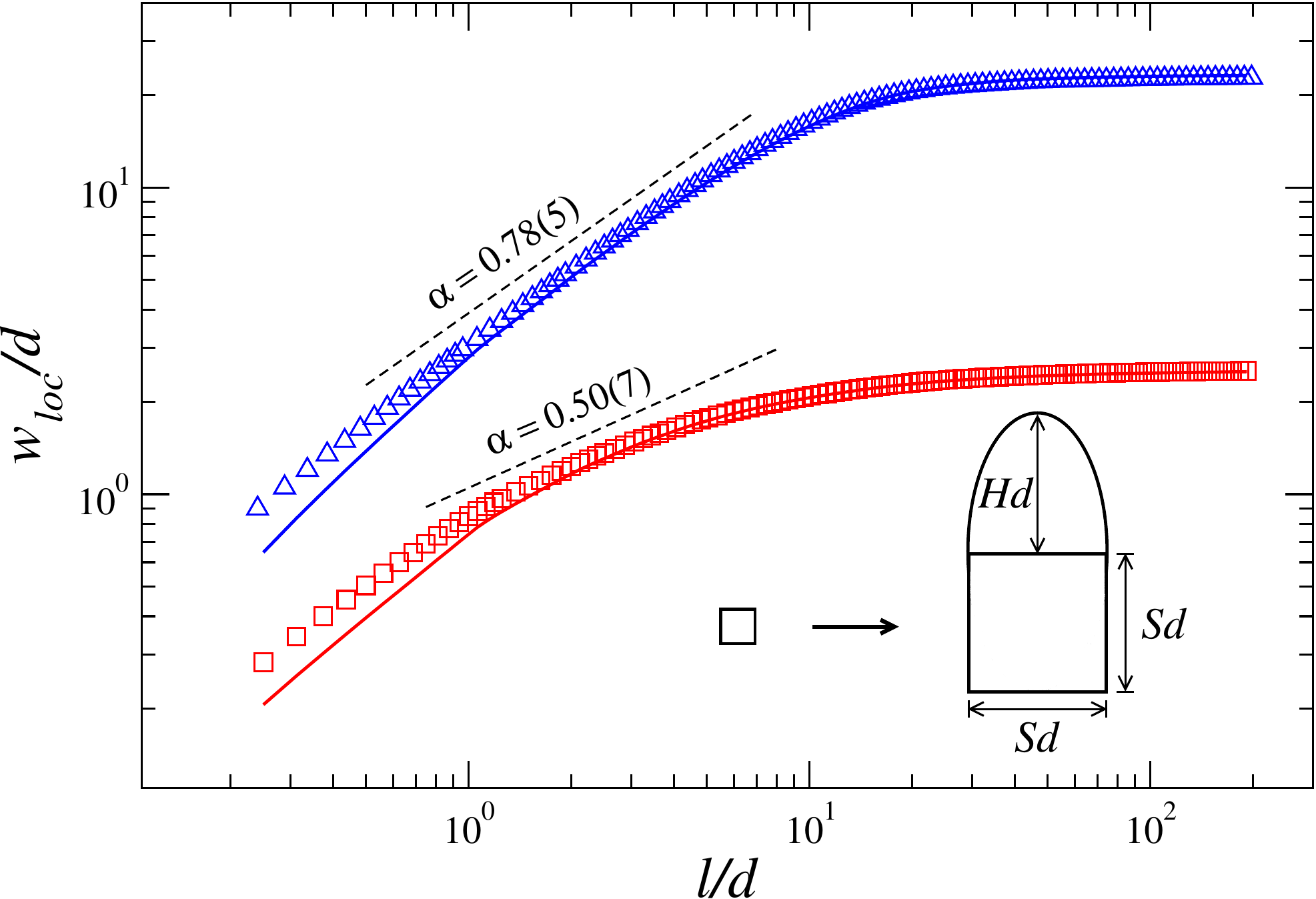}
\caption{(Color online) Local roughness $w_{loc}$ versus the box size $l$, rescaled by the particle size $d$,
for $D=0$ and $H=8$ (red full line), $D=0$ and $H=28$ (red squares), $D=8$ and $H=8$ (blue full line) and $D=8$ and $H=28$ (blue triangles).
Dashed lines have the slopes indicated in the plot. The inset illustrates the model of interface particles with ellipsoidal shape.}
\label{fig5}
\end{figure}

After the deposits are grown, each deposited particle is transformed in a
square of lateral size $Sd$, with integer $S>1$.
Then, a semiellipsoid of horizontal size $Sd$ and height $Hd$ is placed on the top of each
surface square, as illustrated in the inset of Fig. \ref{fig5}.
Here, we consider $S=16$ and values of $H$ ranging from $8$ (circles) to $28$
(semiellipsoids with aspect ratio $3.5$).

Fig. \ref{fig5} shows the local roughness as a function of the box size
for two extremal values of $D$ ($0$ and $8$) and $H$ ($8$ and $28$).

For $H=8$ (circles), a crossover is observed as $l$ crosses the particle size $d$ for
both values of $D$.
For $D=0$, the initial slope is slightly larger than the KPZ one,
$\alpha=0.5$, and evolves to values below that one, until saturating.
The slope indicated in Fig. \ref{fig5}  represents the extremal values in the
range $1\lesssim l/d\lesssim 10$.
For $D=8$, the crossover is less salient, but the slopes are very different from the KPZ one;
they range from $\sim 1$ for $l<d$ to $\approx 0.78$ for $l>d$,
far from saturation. 
The results for $H=28$ (thinnest semiellipsoids) show a smooth crossover effect for $D=0$ 
and a negligible one for $D=8$. In the last case, a slope $0.78(5)$
is obtained in almost two decades of $l$, as illustrated in Fig. \ref{fig5}.

The value of $D$ determines the order of magnitude of the roughness in each box size $l$,
as shown by the large distance of the curves for $D=0$ and $D=8$ in Fig. \ref{fig5}.
This is a consequence of the columnar structure, which increases the value of the roughness
by a factor of order $D^{x/2}$,  as explained in Sec. \ref{interfacegrowth}.
The value of $D$ also determines the scaling for $l>d$, with larger $D$ giving larger effective roughness exponents.

On the other hand, the initial slope ($l < d$) is determined by the particle shape, i. e., by the
value of $H$, as discussed in Ref. \cite{grainshape}. Increasing $H$ corresponds to adding a larger
constant term to the roughness, which is related to the intrinsic width \cite{graos}.
This leads to a decrease of the initial slope.

The presence of a crossover or of a smooth scaling in the $\log{w_{loc}}\times \log{l}$
depends on the combination of these effects. As illustrated in Fig. \ref{fig5}, the particular case
$D=8$, $H=28$, leads to apparently smooth scaling with an exponent much larger than the KPZ value. 
Due to this complex interplay of short range and long
range effects, we are not able to derive a simple relation between the effective roughness exponent
and the parameters $D$ and $H$. However, the trend in all cases is that effective exponents exceed the
KPZ value.

In colloidal particle deposits of Yunker et al \cite{yunker}, the roughness exponents
range from $0.51(5)$ for spherical particles to $0.61(2)$ for the most elongated particles.
The exponents obtained in our model span a much larger range, possibly because all deposited particles
were stretched in the vertical direction and this stretching leads to a nontrivial change
in the scaling (e. g. adding a constant term instead of a multiplicative factor).
However, the trend of the roughness exponent increasing with $\epsilon$ in that experiment
parallels the trend of increase with $D$ in our model.

It is also important to observe that our exponent estimates for
$D=8$ do not obey the scaling relation $z=\alpha/\beta$. However, we stress that they
are effective exponents obtained in a transient regime, with very small film thicknesses, and are
different from the  asymptotic (KPZ) exponent values that do obey that relation.

\section{Asymptotic scaling}
\label{kpz}

The RCA model was shown to belong to the KPZ class in $2+1$ dimensions \cite{perez2005}.
However, as far as we know, no previous work has studied the scaling of that model in $1+1$
dimensions. For this reason, this section is devoted to the study of the asymptotic class
of our deposition model (which is an extension of the RCA model) in $1+1$ dimensions.

\begin{figure}[t]
\includegraphics*[width=8.3cm]{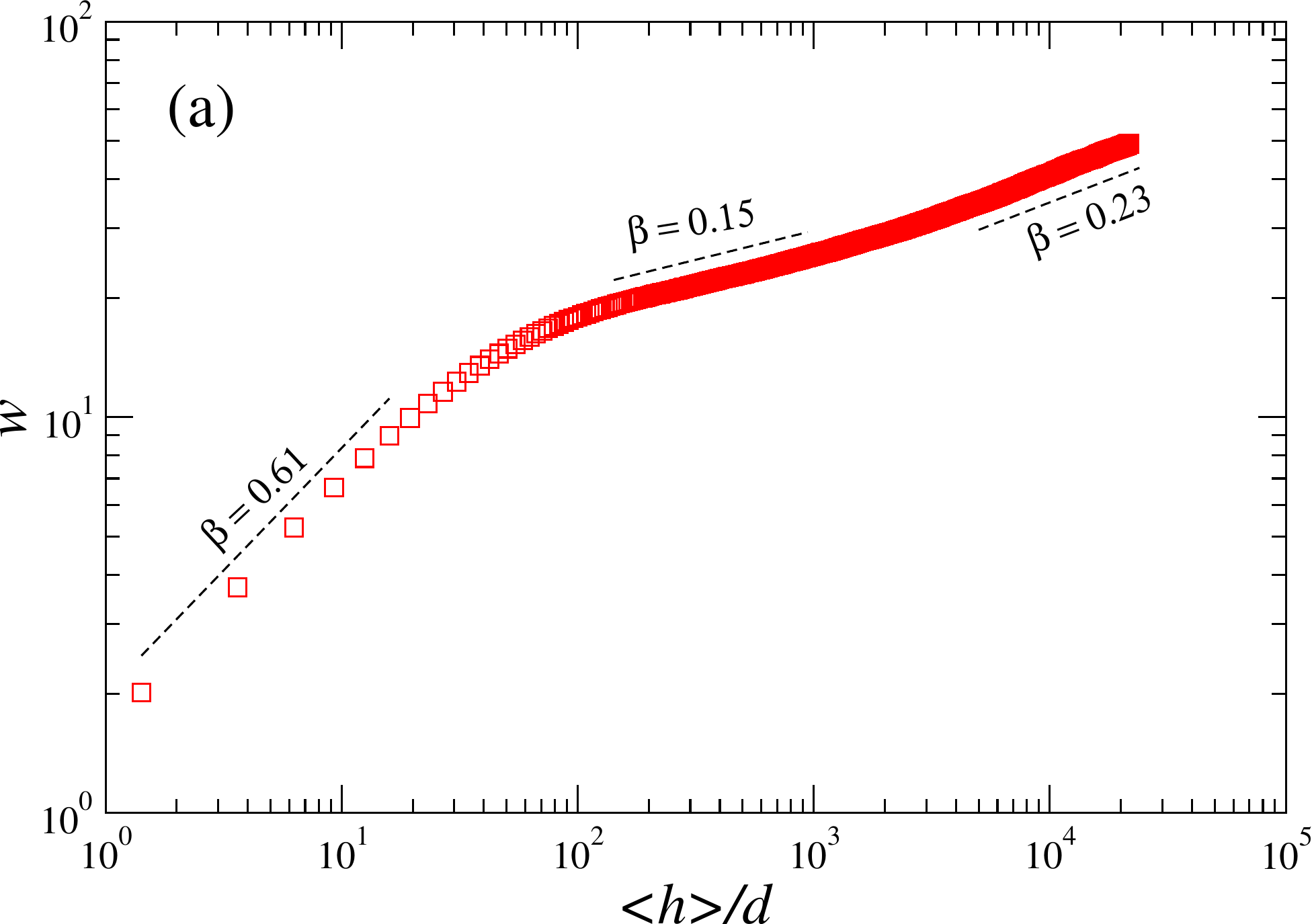}
\includegraphics*[width=8.3cm]{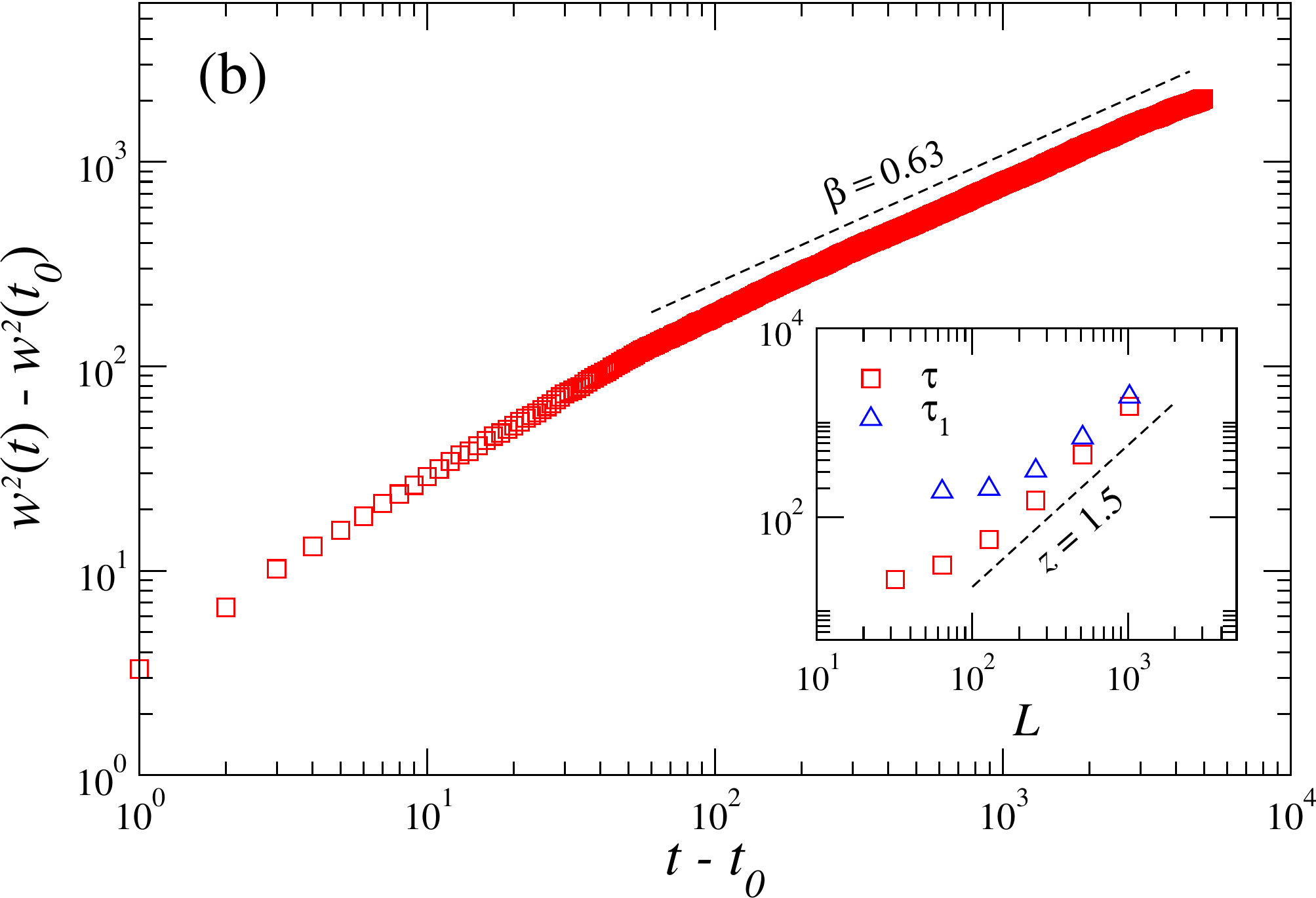}
\caption{(Color online) a) Roughness evolution in time, for $D=4$ and long growth times.
b) Reduced square roughness $w^2\left( t\right) -w^2\left( t_0\right)$ versus reduced time $t - t_0$,
for the same data of (a). Inset shows the characteristic saturation times $\tau$ and $\tau_1$
against the substrate size $L$.}
\label{fig6}
\end{figure}

Fig. \ref{fig6}a shows the time evolution of the global roughness $w$ for the model with $D=4$
in a substrate of size $L=4096$. The maximal time is chosen very far from the saturation regime.
An initial rapid increase of $w$ is observed, with slope near $0.61$, which is related to the
formation of the columnar structure. Subsequently, $w$ slowly increases, with slope near $0.15$
in Fig. \ref{fig6}a. However, this slope is continuously increasing and attains a value near
$0.23$ for $\left\langle h \right\rangle/d > 5000$. These features show that a long crossover is present in the model.
This is not unexpected in a ballistic-like model; indeed, large
scale simulations of BD and related models are necessary to provide asymptotic KPZ exponents with
good accuracy \cite{farnudi}.

Based on the experience with other ballistic-like models, we assume that the square roughness
has the expected asymptotic form proportional to $t^{2\beta}$ with a constant correction
term ${w_I}^2$, which is usually called intrinsic width \cite{intrinsic1,intrinsic2}.
Since estimating this term is typically difficult, our procedure is to analyze the relation
between $w^2(t)-w^2(t_0)$ and $t-t_0$, with a suitable choice of $t_0$. Inspection of
\ref{fig6}a suggests using $t_0=40$, in which $w^2(t_0)=378.6$, since this time slightly exceeds
the transient region. Fig. \ref{fig6}b shows $\log{\left[ w^2\left( t\right) -w^2\left( t_0\right)\right]}$
versus $\log{\left( t-t_0\right)}$, which gives $2\beta\approx 0.63$ in two decades of the
variable $t$, in good agreement with the KPZ value $2\beta = 2/3$.

Additional support for the KPZ scaling is provided by calculation of saturation times of
smaller systems, with sizes between $L=64$ and $L=1024$. Estimating the saturation time from the convergence
of the roughness to the saturation value is a difficult task. However, an alternative is suggested
in Ref. \protect\cite{tau} and is extended here to cases with large intrinsic roughness, as follows.

The Family-Vicsek (FV) scaling relation \cite{FV} [which contains the
relations (\ref{defbeta}) and (\ref{defalpha}) as limiting cases] can be written as
$w^2\left( t,L\right) ={w^2}_{sat}f\left( t/L^z\right)$, where $f$ is a scaling function so that
$f\left( x\right)\to 1$ as $x\to\infty$ and $f\left( x\right)\sim x^\beta$ for $x\ll 1$.
The characteristic time $\tau$ is defined as $w^2\left( \tau_1\right) =k{w^2}_{sat}$, for fixed $L$
and fixed $k<1$. Using the FV relation, we have $\tau\sim L^z$ \cite{tau}.
In the presence of a intrinsic roughness $w^2\left( t_0\right)$, the FV relation may be written as
$w^2\left( t,L\right)-w^2\left( t_0\right)= \left[ {w^2}_{sat} -w^2\left( t_0\right) \right]
g\left[ \left( t-t_0\right)/\tau\right]$,
where the function $g$ has the same asymptotic forms of $f$.
Thus, a characteristic time $\tau_1$ is defined as $w^2\left( \tau_1\right) -w^2\left( t_0\right) =
k\left[ {w^2}_{sat} -w^2\left( t_0\right) \right]$, for fixed $L$ and fixed $k<1$.
For long times, it gives $\tau_1\sim L^z$.

In the inset of Fig. \ref{fig6}b we show $\tau$ and $\tau_1$ as a function of the lattice size $L$.
For small $L$, a slow increase of those characteristic times is observed. However, the largest
values of $L$ suggest a convergence to the KPZ scaling with slope $3/2$. The usual method
of calculating effective exponents from the data in sizes $L$ and $L/2$ is not helpful in 
this case because only the last three data points belong to a scaling region. 

\section{Discussion and conclusion}
\label{conclusion}

We performed a detailed analysis of the short time scaling of height fluctuations of a model
\cite{perez2005}, which
mimics the ``Matthew effect'' recently observed in colloidal particle deposition at the edges of
evaporating drops \cite{yunker}. Increasing the number of horizontal random steps $D$ in the model is
equivalent to increasing the range of attraction between deposited and wandering particles and leads
to increasing effective growth exponents, ranging from $\beta\approx 0.33$ to $\beta\approx 0.68$,
in good agreement with the experimental estimates.
The local roughness was measured accounting for intra-particle scales and provided a broad range of roughness
exponents, also including the estimates of that colloidal particle deposition.
Notwithstanding, the morphology of deposits and height profiles obtained with the model are similar to the ones
shown in Ref. \protect\cite{yunker}.

The large growth and roughness exponents observed here for large $D$ are simple consequences of the
initial columnar growth present in the system. Since this columnar growth is also observed in the experiments
with highly elongated particles $\epsilon$ \cite{yunker}, there is no need to invoke a critical KPZ dynamics
with quenched disorder to explain those features, as previously pointed out by
Nicoli et al. \cite{nicoli}. After this \textit{transient} columnar growth, our model suggests that
a crossover to universal KPZ scaling is possible.

In order to decide between these different interpretations, we propose to measure the dynamic exponent $z$
from the first zero or minima of the autocorrelation function. For small $D$,
this function have the behavior
expected for self-affine interfaces and the corresponding dynamic exponent agrees with the KPZ value $z=1.5$.
However, for large $D$, the autocorrelation function has an oscillatory behavior due to the
columnar structures and provides much larger estimates of $z$ (far from
the QKPZ value $z \approx 1$). This proposal may also be very important for future works to
decide between a true dynamic scaling and possible transient behaviors, particularly because the scaling
law $z=\alpha/\beta$ does not hold for the latter.

\section*{Acknowledgments}

The authors thank R. Cuerno for helpful discussions. This work was supported by FAPEMIG, FAPERJ, and CNPq (Brazilian agencies).

\section*{References}

\bibliography{collpartdep}

\end{document}